# Hysteresis and Compensation behaviors of spin-1 Blume–Capel model on the hexagonal Ising nanowire

**Yusuf Kocakaplan**[1, *] **and Ersin Kantar**[2]

[1]Graduate School of Natural and Applied Sciences, Erciyes University, 38039 Kayseri, Turkey
[2]Department of Physics, Erciyes University, 38039 Kayseri, Turkey

**Abstract**

An effective-field theory with correlations has been used to study the hysteresis, susceptibility and compensation behaviors of the spin-1 hexagonal Ising nanowire (HIN) with core-shell structure. The effects of the temperature, crystal field, core-shell interfacial coupling and shell surface coupling are investigated on hysteresis and compensation behaviors, in detail. When the core-shell interfacial coupling is weak, the double and triple hysteresis loops can be seen in the system. The hysteresis loops have different coercive field points that the susceptibility make peak at these points. Moreover, we observed that the system displays the Q-, R-, S-, P-, N- and W-types of compensation behaviors according to Neél classification nomenclature. Finally, the obtaining results are compared with some experimental and theoretical results and found a good agreement.

*Keywords:* Hexagonal Ising nanowire; Effective-field theory; Hysteresis behavior; Susceptibility; Compensation behavior

## 1. Introduction

The spin-1 Ising model in the presence of crystal-field (D) which called as Blume–Capel (BC) model is one of the most actively studied higher spin Ising models in statistical physics due to the both theoretical importance and its application to different systems [1-9]. BC model has been studied with many different techniques, such as mean field approximation (MFA) [10], effective field theory (EFT) [11-13], Bethe approximation [14], Monte Carlo simulation

[*] Corresponding author.
Tel: + 90 352 2076666 # 33139
*E-mail address:* yusufkocakaplan@gmail.com (Y. Kocakaplan)

(MCs) [15], renormalization group theory [16] and the cluster variational approximation [17, 18], and it was found that it exhibits very interesting and rich critical phenomena.

On the other hand, from theoretical point of view, many studies of magnetic properties of nanostructured materials (nanowire, nanotube and nanoparticle etc.) have been investigated by a variety of techniques in the equilibrium statistical physics. The first reason is that they have a wide possibility of technological applications, such as ultrahigh-density magnetic storage devices [19], sensors [20], permanent magnets [21], and medical applications [22]. The second one is that many rich phenomena are observed in these systems. Recently, magnetic properties of nanoparticles have been studied within various Ising systems consisting of core-shell structures. Kaneyoshi investigated by using the EFT with correlations phase diagrams [23], magnetizations [24], the possibility of a compensation point [25] and reentrant phenemona [26] of the spin-1/2 Ising nanoparticles system. Kocakaplan and Kantar studied the thermal and magnetic properties of the spin-1/2 hexagonal Ising nanowire (HIN) [27]. Boughazi *et al*. [28] obtained the phase diagrams and magnetic properties of cylindrical spin-1 Ising nanowire via the MC simulations. Kocakaplan *et al.* [29] studied hysteresis loops and compensation behavior of cylindrical transverse spin-1 Ising nanowire by the EFT based on a probability distribution technique. Canko *et al.* [30] examined thermodynamic properties of spin-1 Ising nanotube. Magoussi *et al.* are examined the hysteresis behaviors [31], and phase diagrams and magnetic properties [32] of a spin-1 Ising nanotube. Furthermore, Jiang *et al.* [33] studied spin-3/2 core and a hexagonal ring spin-5/2 shell by the EFT with self-spin correlations. Kantar and Kocakaplan [34] presented the phase diagrams and compensation behaviors HIN system with spin-1/2 core and spin-1 shell structure.

In the present work, we studied the hysteresis and compensation behaviors of the spin-1 hexagonal Ising nanowire with core-shell structure via the EFT with correlations. As far as we know, spin-1 Ising model has not yet been investigated on hexagonal type Ising nanowire.

The paper is arranged as follows. In Section 2, we give the model and present the formalism of the model in the EFT with correlation. The detailed numerical results and discussions are presented in Section 3. Finally Section 4 is devoted to a summary and a brief conclusion.

## 2. Model and formulation

The Hamiltonian of the hexagonal Ising nanowire (HIN) includes nearest neighbors interactions and the crystal field is given as:

$$H = -J_S \sum_{\langle ij \rangle} S_i S_j - J_C \sum_{\langle mn \rangle} \sigma_m \sigma_n - J_1 \sum_{\langle im \rangle} S_i \sigma_m - D \left( \sum_i S_i^2 + \sum_j S_j^2 + \sum_m \sigma_m^2 + \sum_n \sigma_n^2 \right)$$
$$- h \left( \sum_i S_i + \sum_j S_j + \sum_m \sigma_m + \sum_n \sigma_n \right) \quad (1)$$

where $\sigma = \pm 1$ and $S = \pm 1, 0$. The $J_S$, $J_C$ and $J_1$ are the exchange interaction parameters between the two nearest-neighbor magnetic particles at the shell surface, core and between shell surface and core, respectively (see Fig. 1). $D$ is Hamiltonian parameter and stand for the single-ion anisotropy (i.e. crystal field). The surface exchange interaction and core-shell interfacial coupling are often defined as $J_S = J_C(1+\Delta_S)$ and $r = J_1/J_C$ to clarify the effects of the shell surface and core-shell interactions on the physical properties in the nanosystem, respectively.

Within the framework of the EFT with correlations, one can easily find the magnetizations $m_S$, $m_C$ and the quadruple moment $q_s$ as coupled equations, for the mixed spin-1 HIN system as follows:

$$m_S = \left[1 + m_S \sinh(J_S \nabla) + m_S^2 (\cosh(J_S \nabla) - 1)\right]^4 \left[1 + m_C \sinh(J_1 \nabla) + m_C^2 (\cosh(J_1 \nabla) - 1)\right] F(x)\big|_{x=0}, \quad (2a)$$

$$m_C = \left[1 + m_C \sinh(J_C \nabla) + m_C^2 (\cosh(J_C \nabla) - 1)\right]^2 \left[1 + m_S \sinh(J_1 \nabla) + m_S^2 (\cosh(J_1 \nabla) - 1)\right]^6 F(x)\big|_{x=0}, \quad (2b)$$

$$q_S = \left[1 + m_S \sinh(J_S \nabla) + m_S^2 (\cosh(J_S \nabla) - 1)\right]^4 \left[1 + m_C \sinh(J_1 \nabla) + m_C^2 (\cosh(J_1 \nabla) - 1)\right] G(x)\big|_{x=0}, \quad (2c)$$

$$q_C = \left[1 + m_C \sinh(J_C \nabla) + m_C^2 (\cosh(J_C \nabla) - 1)\right]^2 \left[1 + m_S \sinh(J_1 \nabla) + m_S^2 (\cosh(J_1 \nabla) - 1)\right]^6 G(x)\big|_{x=0}, \quad (2d)$$

where $\nabla = \partial/\partial x$ is the differential operator. The functions $F(x)$ and $G(x)$ are defined as

$$F(x) = \frac{2\sinh[\beta(x+h)]}{\exp(-\beta D) + 2\cosh[\beta(x+h)]} \quad (3a)$$

$$G(x) = \frac{2\cosh[\beta(x+h)]}{\exp(-\beta D) + 2\cosh[\beta(x+h)]} \quad (3b)$$

Here, $\beta = 1/k_B T$, T is the absolute temperature and $k_B$ is the Boltzmann constant. By using the definitions of the order parameters in Eqs. (2a)-(2b), the total ($m_T$) magnetizations of per site can be defined as $m_T = 1/7(6m_S + m_C)$.

In order to obtained susceptibilities of the system, we differentiated magnetizations respect to h as following equation:

$$\chi_\alpha = \lim_{\to 0}\left(\frac{\partial m_\alpha}{\partial h}\right) \tag{4}$$

where, $\alpha$ = C and S. By using of Eqs. (2) and (4), we can easily obtain the $\chi_C$ and $\chi_S$ susceptibilities as follow:

$$\chi_C = a_1 \chi_C + a_2 \chi_S + a_3 \frac{\partial F(x)}{\partial h}, \tag{5a}$$

$$\chi_S = b_1 \chi_S + b_2 \chi_C + b_3 \frac{\partial F(x)}{\partial h}. \tag{5b}$$

Here, $a_i$ and $b_i$ (i=1, 2 and 3) coefficients have complicated and long expressions, hence they will not give. The total susceptibility of per site can be obtain via $\chi_T = 1/7(\chi_C + 6\chi_S)$.

Solving these equations, we can get the numerical results of the spin-1 HIN system. We will perform these results in the next Section.

## 3. Numerical results and discussions

The hysteresis, susceptibility and compensation behaviors of the spin-1 HIN with core-shell structure have investigated and discussed for selected values of the system parameters in the following.

### 3.1. *Hysteresis behaviors of spin-1 HIN system*

In this part, the effects of the temperature, crystal field core-shell interfacial coupling and shell surface coupling on the hysteresis behaviors have been presented, in detail.

### 3.1.1. The effect of the temperature on the hysteresis and susceptibility behaviors

Fig. 2 is obtained for selected four typical temperature values, namely T = 0.5, 1.25, 2.0 and 2.5, in the case of r = -1.0, $\Delta_S$ = -0.99 and D = 0.0 fixed values to investigate the temperature dependence of the hysteresis and susceptibility behaviors of the spin-1 HIN system. In Fig. 2(a), the hysteresis consists triple loops. The susceptibility make six peaks at coercive field points. With the increase of temperature, the triple hysteresis loop turn to single hysteresis loop as seen in Fig. 2(b)-2(c) and when the temperature approaches its critical value, the single hysteresis loop becomes narrower and thinner. The reason of this fact is that all magnetic moments fluctuate with a relaxation time shorter than the measuring time at high temperature values. Thus, magnetizations become very small and from this reason the external magnetic field need to reverse magnetic moments will be less [35, 36]. If the temperature grows stronger, the single hysteresis loop disappears in the case of T > $T_C$ ($T_C$= 2.4) as seen in Fig. 2(d). Moreover, the hysteresis loops area decrease as the temperature increases. This fact is also understand from the susceptibility peak turn to reaches a single value in both directions of external magnetic field. The physical means of this fact is that while at low temperatures the system becomes hard magnet, with the increase of the temperature the hard magnet turn to soft magnet. These results are consistent with some experimental results [35-41].

In the case of r = 0.01, $\Delta_S$ = 0.0 and D = 0.0 fixed values, Fig. 3 is obtained for selected four typical temperature values, namely T = 0.5, 1.5, 2.25 and 2.7, to present the temperature dependence of the hysteresis and susceptibility behaviors for small interfacial coupling value. If the interaction between core and shell to be weak, then the correlation between core and shell will decrease and the system will lose its magnetizations. In Fig. 3(a), we can see that there is only one hysteresis loop that have a rectangular shape owing to the fact that magnetization rapidly reaches its saturation value. The hysteresis loops have four coercive field points that the susceptibility make four peaks at these points as seen in Fig. 3(a). With increasing of the temperature, the rectangular shape of the hysteresis loops start to turn round due to fluctuations. Also, the hysteresis loop becomes narrower as the temperature increases below the transition temperature and the hysteresis loop disappear above the transition temperature ($T_C$ = 2.65). Above transition temperature, the susceptibility displays a single peak as seen in Fig. 3(d).

### 3.1.2. The effect of crystal field on the hysteresis and susceptibility behaviors

In order to discuss the influence of the crystal field on the hysteresis and susceptibility, Fig. 4 is plotted for r = 0.01, $\Delta_S$ = 0.0 and T = 0.5 fixed values, and D = -0.25, -2.0, -2.5 and -5.0 crystal field values. At first, we can see that there is only one hysteresis loop and four coercive field points in Fig. 6(a) for D = -0.25. At these points, the susceptibility make four peaks as seen in Fig. 4(a). With decreasing of the crystal field, hysteresis loop area is narrowing. This fact is clearly seen in Fig. 4(b). As from D = -2.5 value, the single loop start turn to the double loop and to clearly see double loop we plotted magnetizations curves for D = -5.0. As we clearly see from Figs. 4(d), while the crystal field decreases, the hysteresis loops and the susceptibility peaks are diverge from each other for a particular range value of the external magnetic field.

### 3.1.3. The effect of core-shell interfacial coupling on the hysteresis and susceptibility behaviors

In Fig. 5, we show the core-shell interfacial coupling dependence of the hysteresis loops of the HIN system at T = 0.5, D = 0.0 for $\Delta_S$ = -0.5, and r = -0.01, -1.0 and -3.0. We can see that when the antiferromagnetic core-shell interfacial coupling is small, the hysteresis consists one loop as seen in Fig. 5(a). One can see that with the absolute value of r increases, the hysteresis behavior changes from one loop to triple loops. This fact is clearly seen from the Fig. 5(b). Moreover, while the susceptibility make four peaks at the four coercive field points in Fig. 5(a), it makes six peaks at the six coercive field points at the Figs. 5(b) and 5(c). Similar behaviors of the hysteresis loops for different values core-shell interfacial couplings have been found in Ref. [27].

### 3.2. The total magnetization behavior of spin-1 HIN system

Fig. 6(a) displays the effect of the crystal field on the total magnetization behavior. Fig. 6(a) is obtained for r = -0.2, $\Delta_S$ = -0.9 fixed values and D = -0.65, -0.9 and -0.3. In this figure, the W-, S- and N-type of compensation behaviors are obtained for D = -0.65, -0.9 and -0.3 values, respectively. Fig. 6(b) is obtained for r = -0.1, D = -0.6 fixed values and for $\Delta_S$ = -0.99, -0.73, and -0.5 values, and the effect of the surface shell coupling on the total magnetization behavior is investigated. The total magnetization curves labeled with $\Delta_S$ = -0.99, -0.73, and -

0.5 values display the S-, N- and Q-type behaviors. It can be easily seen from Fig. 6(b) that phase transition temperature is growing with the increase of the $\Delta_S$ values.

### 3.3. *The Compensation types of spin-1 HIN system*

As known, the existence of the compensation temperature in a magnetic nanoparticle has important applications in the field of thermo-magnetic recording. In this purpose, we also studied the temperature variation of the total magnetization for various values of physical parameters of the system to obtain the compensation temperature and determine compensation types. Fig. 7(a) shows the Q-type behaviors for the curve labeled r = -1.0, $\Delta_S$ = 0.0 and D = 0.0. The R-type behavior is obtained in Fig. 7(b) for r = -1.0, $\Delta_S$ = -0.5 and D = -0.75. For r = -2.0, $\Delta_S$ = -0.99 and D = 0.0 values, the S-type behaviors is obtained as seen in Fig. 7(c). Fig. 7(d) indicates the P-type behaviors for r = -0.1, $\Delta_S$ = 1.0 and D = 0.0 values. For r = -0.25, $\Delta_S$ = -0.75 and D = -1.0 values, the N-type behaviors have observed as seen in Fig. 7(e). Finally, the W-type behavior and two compensation points have obtained in Fig. 7(f) for r = -0.2, $\Delta_S$ = -0.9 and D = -0.75. The same compensations behaviors classified in the Néel theory [42]. It is also worth noting that recently the Q-, R-, S- and N-type [34] and the Q-, R-, N-, M-, P-, and S- type [29] behaviors have been obtained in the mixed Ising nanoparticles and mixed hexagonal Ising nanowire systems, respectively.

## 4. Summary and Conclusion

In this study, we studied the hysteresis, susceptibility and compensation behaviors of the spin-1 hexagonal Ising nanowire with core-shell structure in the existence of the crystal field by means of the EFT with correlations. The effects of the temperature, crystal field, core-shell interfacial coupling and shell surface coupling are investigated on hysteresis, susceptibility and compensation behaviors. Beside the single hysteresis loop, the double and triple hysteresis loops have observed in the system for the low values of core-shell interfacial coupling. Our results show that the hysteresis loops consist of various coercive field points that the susceptibility make peak at these points. The Q-, R-, S-, P-, N- and W-types of compensation behaviors have obtained in the system according to the Néel classification nomenclature. Finally, we hope that our detailed theoretical investigation may stimulate further researches to study magnetic properties of nanoparticles systems, and also will motivate experimentalists to investigate the compensation behavior in real nano material systems.

**List of the Figure Captions**

**Fig. 1.** (Color online) Schematic representation of hexagonal Ising nanowire. The blue and red spheres indicate magnetic atoms at the surface shell and core, respectively.

**Fig. 2.** (Color online) Hysteresis and susceptibility behaviors of the spin-1 HIN system for r = -1.0, $\Delta_S$ = -0.99, D = 0.0 and for various values of temperatures.
   **(a)** T = 0.5; **(b)** T = 1.25; **(c)** T = 2.0; **(d)** T = 2.5.

**Fig. 3.** (Color online) Same as Fig. (2), but for r = 0.01, $\Delta S$ = 0.0, D = 1.0, and
   **(a)** T = 0.5; **(b)** T = 1.5; **(c)** T = 2.25; **(d)** T = 2.7.

**Fig. 4.** (Color online) Same as Fig. (2), but for r = 0.01, $\Delta S$ = 0.0, T = 0.5, and
   **(a)** D = -0.25; **(b)** D = -2.0; **(c)** D = -2.5; **(d)** D = -5.0.

**Fig. 5.** (Color online) Same as Fig. (2), but for $\Delta_S$ = -0.5, D = 0.0, T = 0.5, and
   **(a)** r = -0.01; **(b)** r = -1.0; **(c)** D = -3.0.

**Fig. 6.** (Color online) The thermal variations of the total magnetization for
   **(a)** For r = -0.2, $\Delta_S$ = -0.9, and D = -0.65, D = -0.9 and D = -0.3.

**(b)** For r = -0.1, D = -0.6, and $\Delta_S$ = -0.99, $\Delta_S$ = -0.73 and $\Delta_S$ = -0.5.

**Fig. 9.** The type of compensation behaviors for
- **(a)** r = -1.0, $\Delta_S$ = 0.0 and D = 0.0.
- **(b)** r = -1.0, $\Delta_S$ = -0.5 and D = -0.75.
- **(c)** r = -2.0, $\Delta_S$ = 0.99 and D = 0.0.
- **(d)** r = -0.1, $\Delta_S$ = 1.0 and D = 0.0.
- **(e)** r = -0.25, $\Delta_S$ = -0.75 and D = -1.0.
- **(f)** r = 0.2, $\Delta_S$ = -0.9 and D = -0.6.

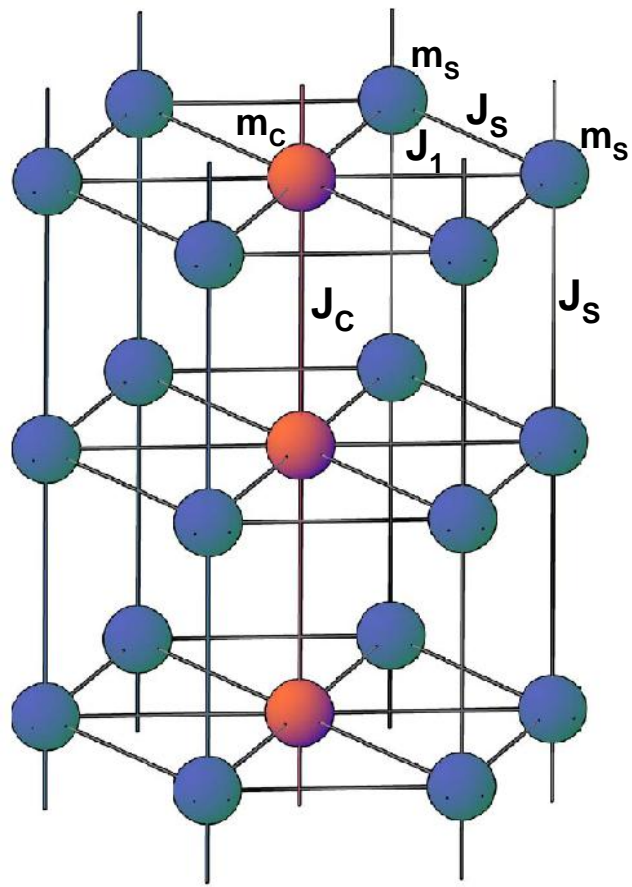

Fig. 1

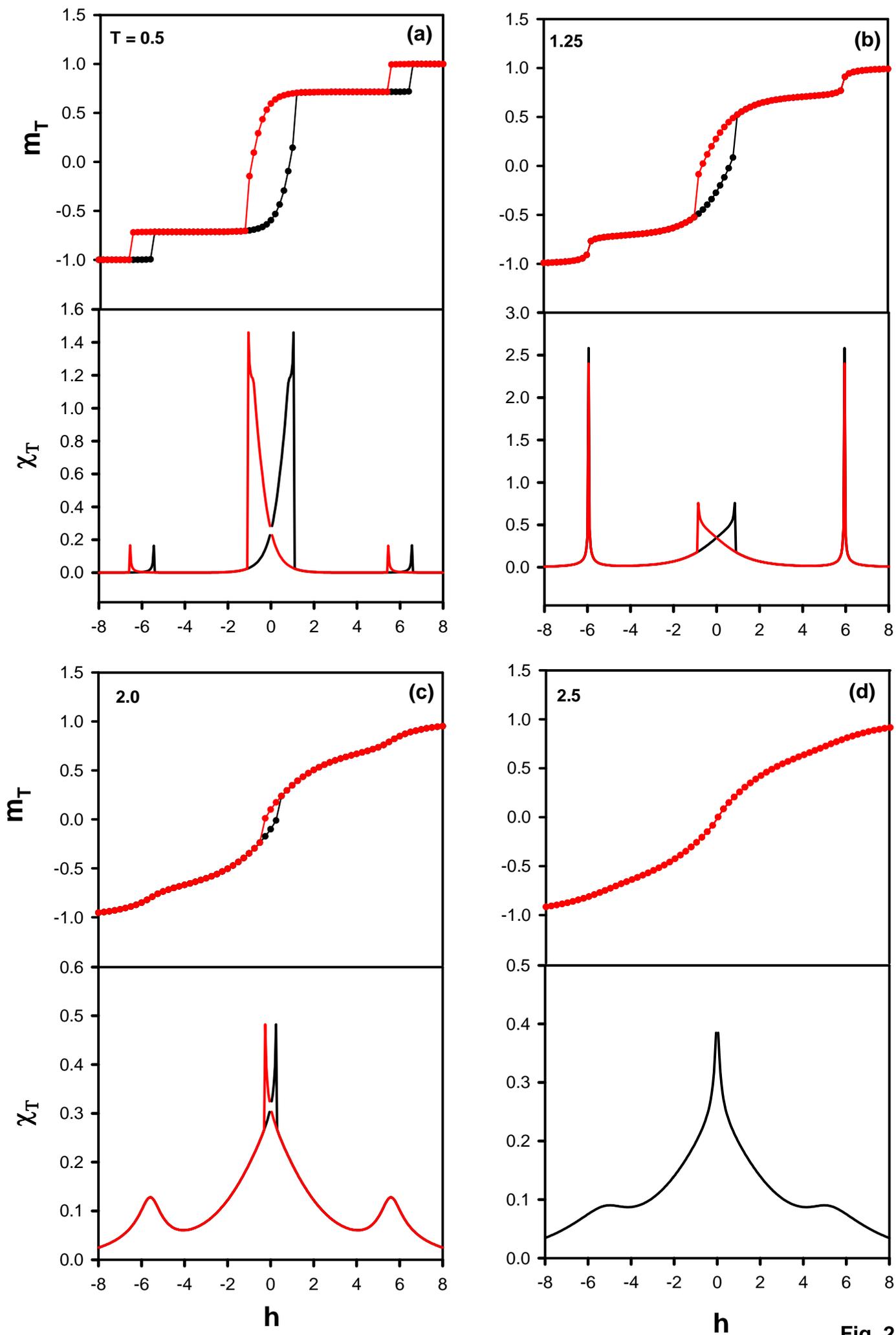

Fig. 2

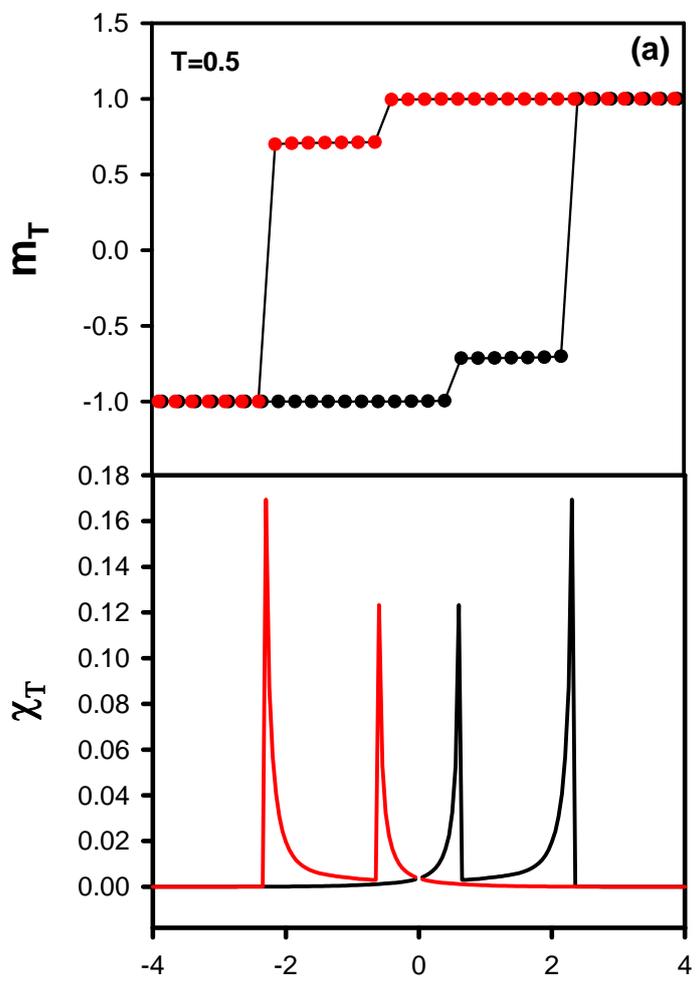
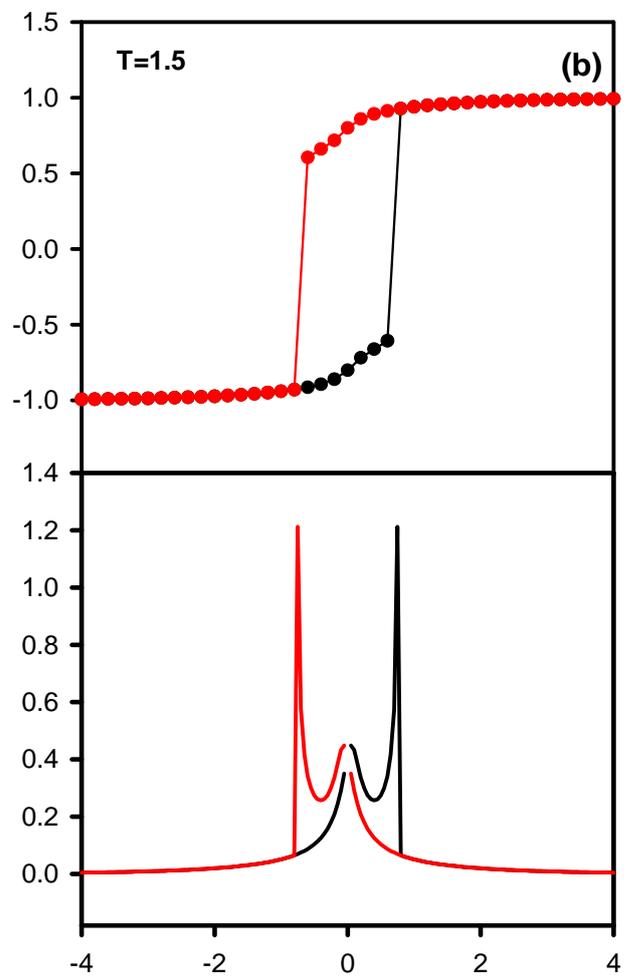
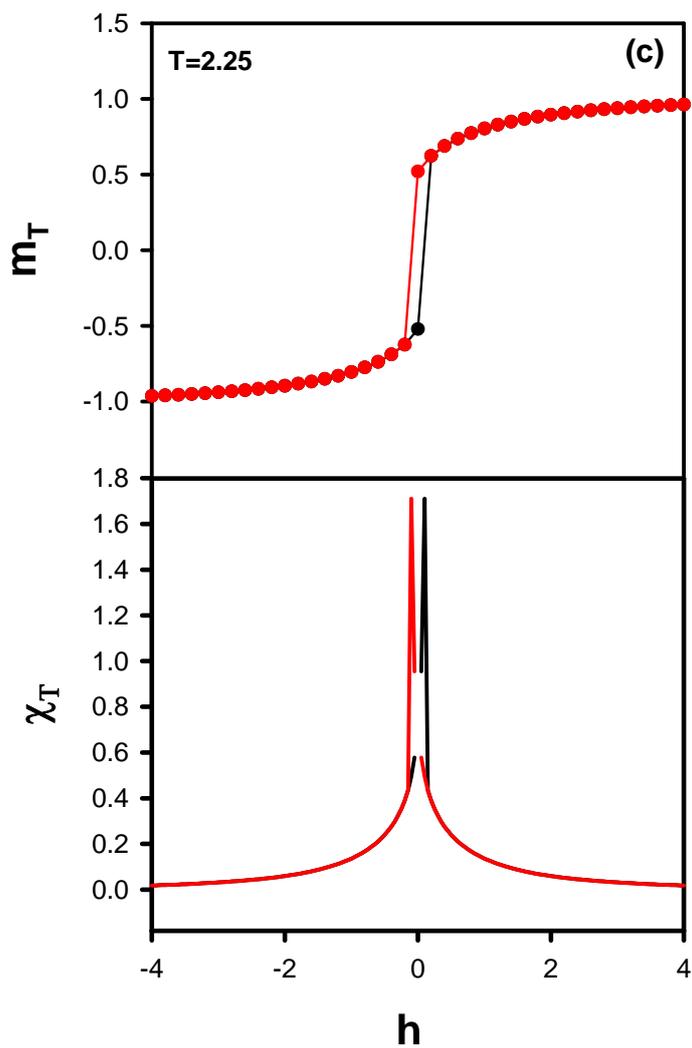
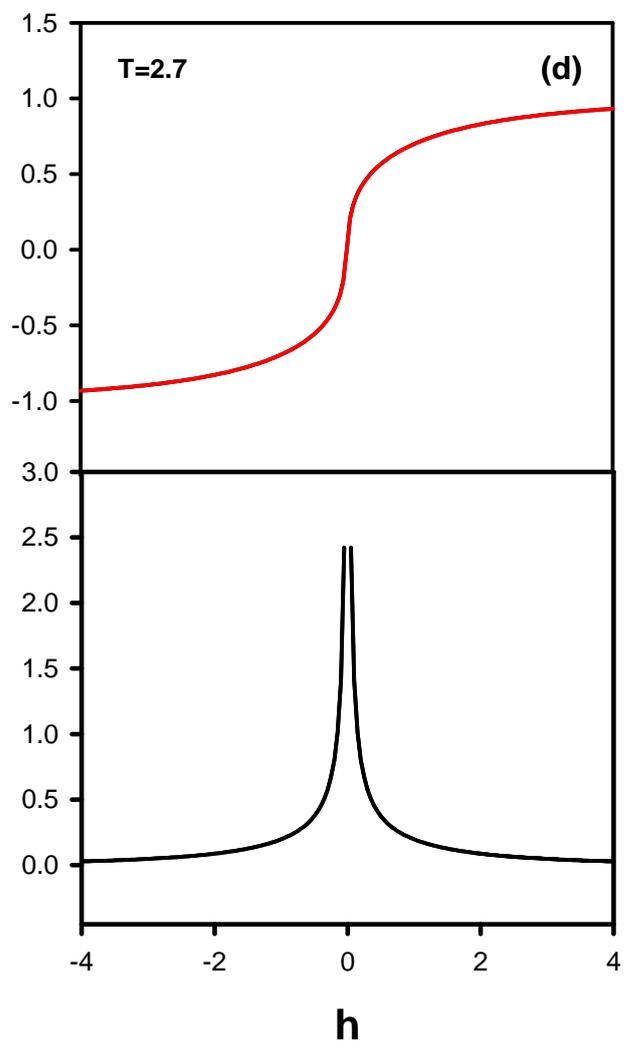

Fig. 3

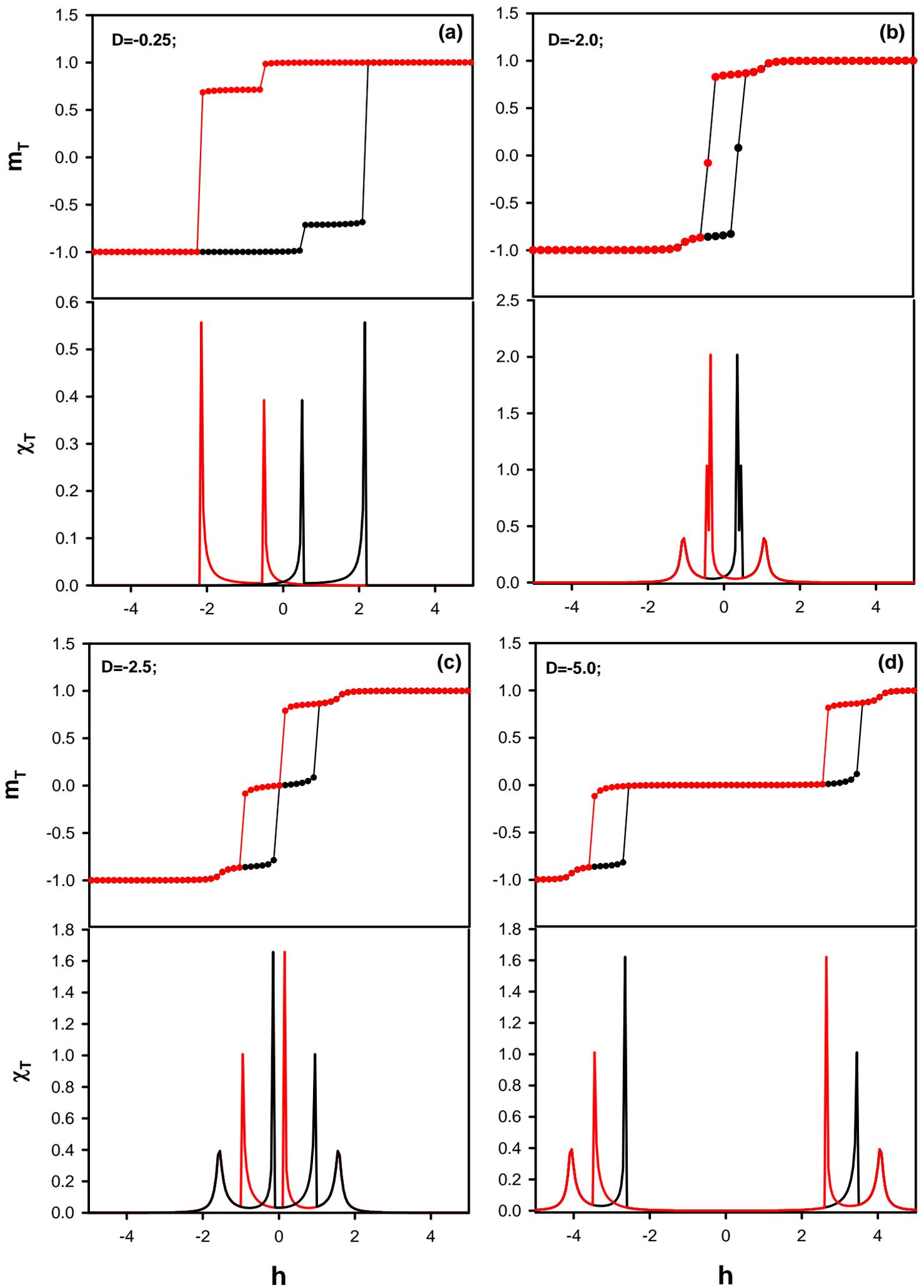

Fig. 4

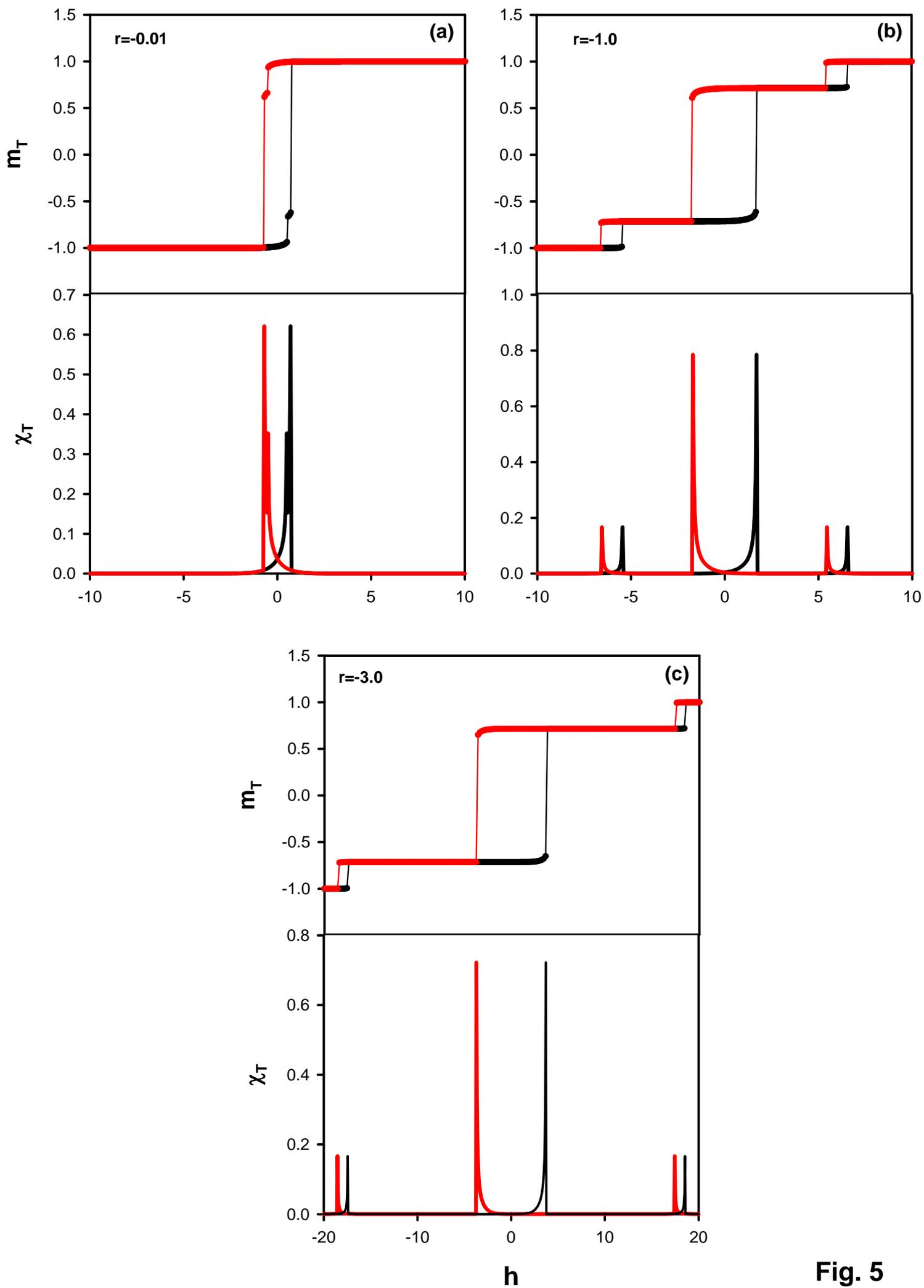

Fig. 5

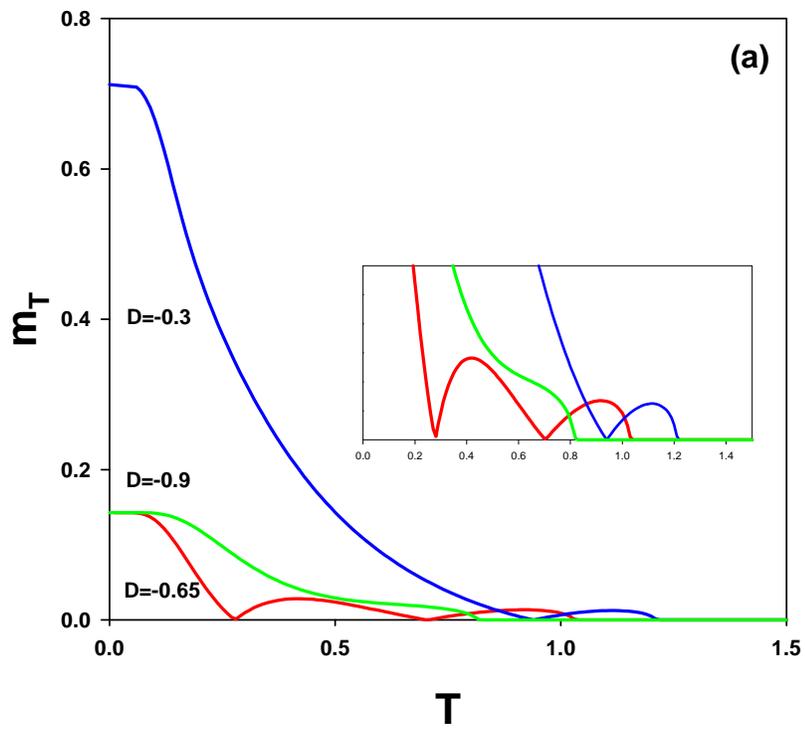

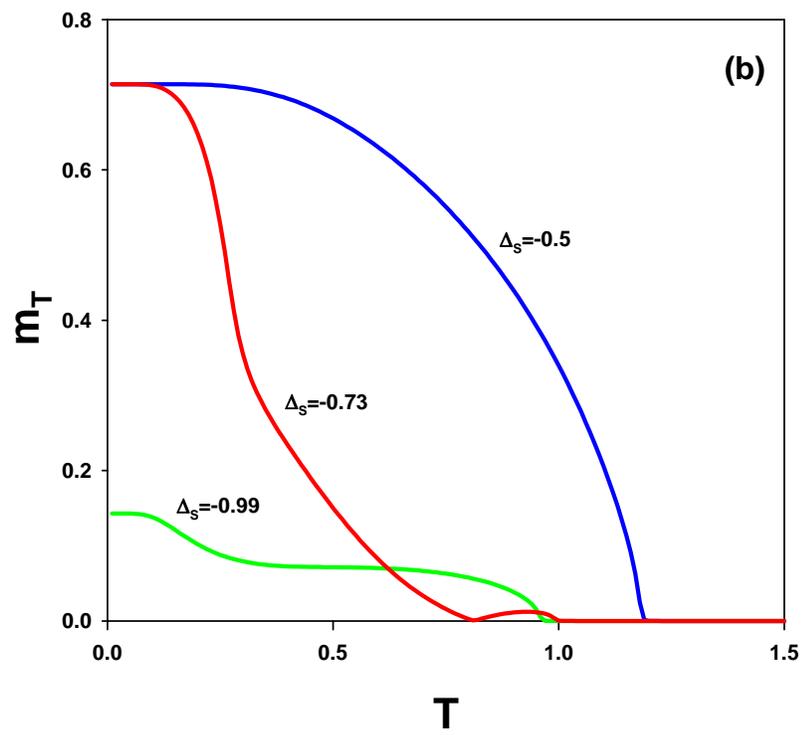

Fig. 6

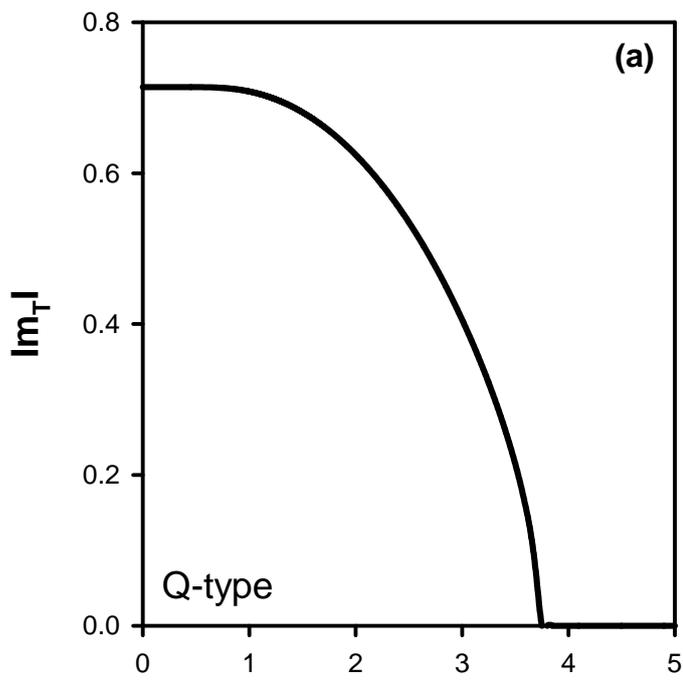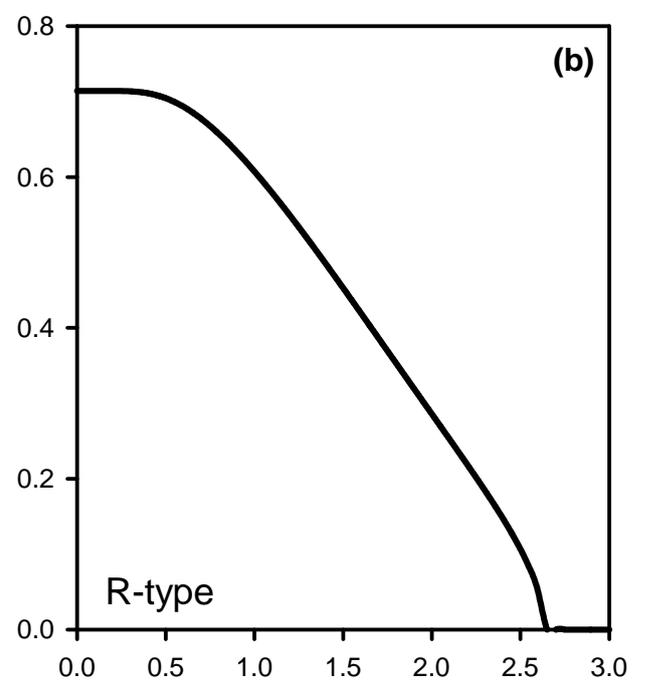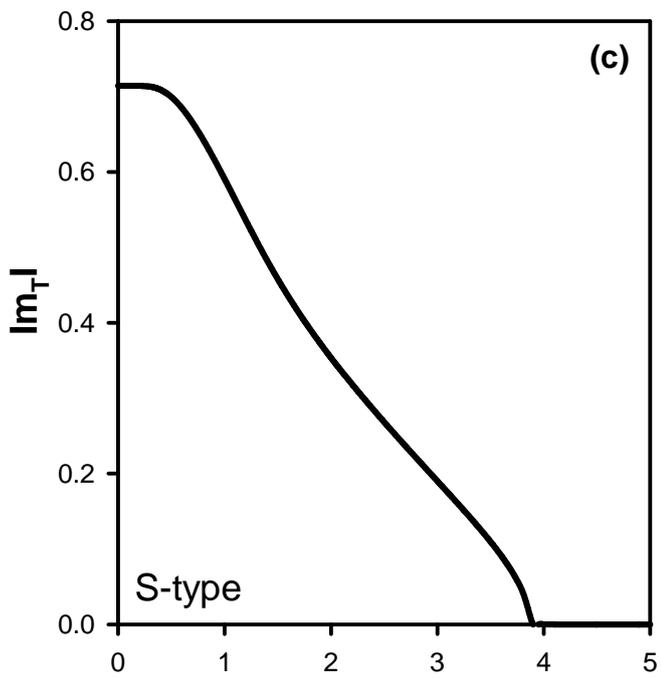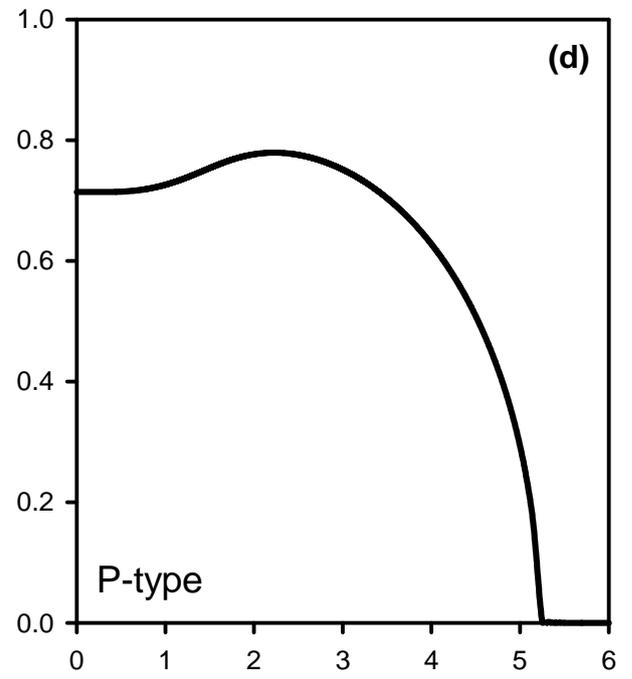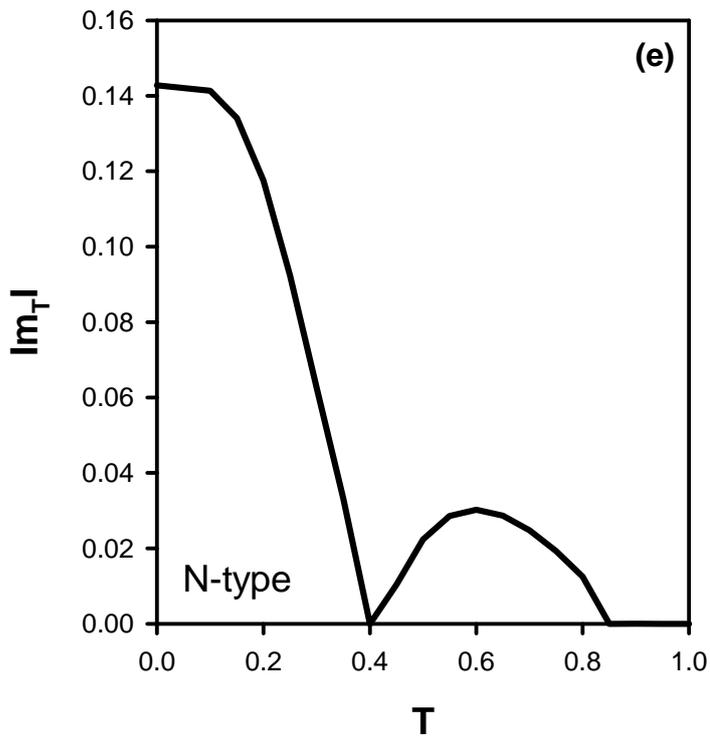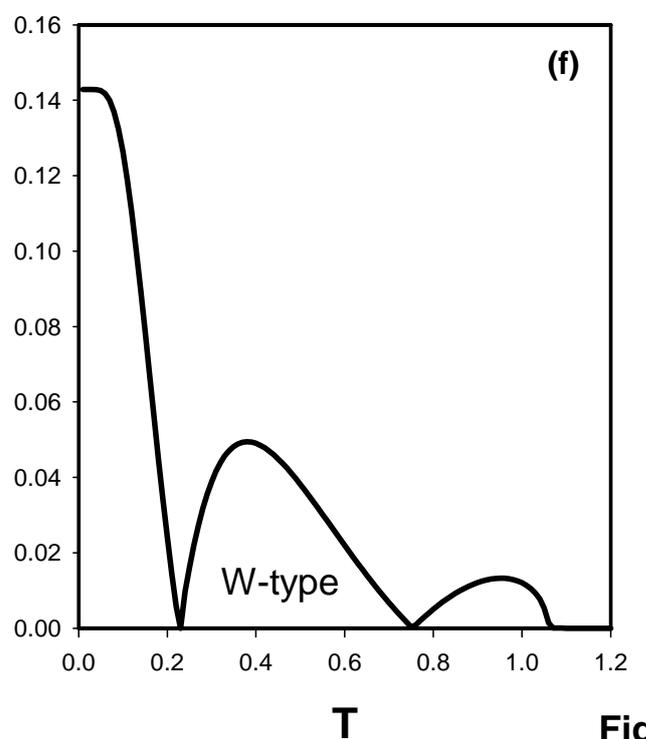

Fig. 7